%% file: paper.tex
\pdfoutput=1
\documentclass[11pt]{article}
\usepackage[utf8]{inputenc}
\usepackage[margin=1in]{geometry}
\usepackage{times} \usepackage{amsmath} \usepackage{amssymb} \usepackage{amsthm}
\usepackage{mathtools} \usepackage{microtype}
\usepackage{graphicx} \usepackage{tikz} \usepackage{booktabs} \usepackage{array}
\usepackage{float}
\usepackage[round,authoryear]{natbib}
\usepackage{algorithm}
\usepackage{xcolor} \usepackage{aliascnt}
\usepackage[colorlinks=true,linkcolor=blue,citecolor=blue,urlcolor=blue]{hyperref}
\usepackage[nameinlink,capitalise,noabbrev]{cleveref}
\hypersetup{hypertexnames=false}
\usepackage{algpseudocode}
\let\cite\citep

\input{bib_macros}

\usetikzlibrary{arrows.meta,positioning,calc,matrix,fit,backgrounds}

\theoremstyle{plain}
\newtheorem{theorem}{Theorem}[section]
\newaliascnt{lemma}{theorem} \newtheorem{lemma}[lemma]{Lemma} \aliascntresetthe{lemma}
\newaliascnt{proposition}{theorem}  \aliascntresetthe{proposition}
\newaliascnt{corollary}{theorem}  \aliascntresetthe{corollary}
\newaliascnt{conjecture}{theorem}  \aliascntresetthe{conjecture}
\newaliascnt{claim}{theorem}  \aliascntresetthe{claim}
\newaliascnt{observation}{theorem}  \aliascntresetthe{observation}
\newaliascnt{fact}{theorem}  \aliascntresetthe{fact}
\theoremstyle{definition}
\newaliascnt{definition}{theorem} \newtheorem{definition}[definition]{Definition} \aliascntresetthe{definition}
\newaliascnt{problem}{theorem} \newtheorem{problem}[problem]{Problem} \aliascntresetthe{problem}
\newaliascnt{assumption}{theorem}  \aliascntresetthe{assumption}
\newaliascnt{remark}{theorem} \newtheorem{remark}[remark]{Remark} \aliascntresetthe{remark}

\crefname{problem}{problem}{problems}
\Crefname{problem}{Problem}{Problems}
\newcommand{\I}{\mathcal I}
\newcommand{\E}{\mathbb E}
\newcommand{\Prb}{\mathbb P}

\newcommand{\OPT}{\operatorname{OPT}}
\newcommand{\ALG}{\operatorname{ALG}}
\newcommand{\eps}{\varepsilon}

\newcommand{\wh}{\widehat}
\newcommand{\vhat}{\widehat v}
\newcommand{\xhat}{\widehat x}
\newcommand{\Ohat}{\widehat O}
\newcommand{\Mhat}{\widehat M}
\newcommand{\rk}{\operatorname{rk}}
\newcommand{\one}{\mathbf 1}

\newcommand{\bv}[1]{\mathbf{#1}}

\newcommand{\hv}{\widehat{\bv v}}
\newcommand{\defeq}{\mathrel{:=}}
\newcommand{\Base}{\mathcal A_{\mathrm{base}}}

\newcommand{\whx}{\bv{\widehat x}}

\usepackage{enumitem}
\setlist[itemize]{
  itemsep=2pt,
  parsep=0pt,
  topsep=3pt,
  partopsep=0pt
}

\begin{document}

\title{Constrained Online Selection with Predictions:\\ Matroid Secretary, Knapsack, and Matching}
\author{%
Hedyeh Beyhaghi\\
\small University of Massachusetts Amherst\\
\small \texttt{hbeyhaghi@umass.edu} \and Tao Zixu He\\
\small University of Massachusetts Amherst\\
\small \texttt{taohe@umass.edu} \and Cameron Musco\\
\small University of Massachusetts Amherst\\
\small \texttt{cmusco@umass.edu}}
\date{}
\maketitle

\begin{abstract}

We study learning-augmented algorithms for constrained online selection problems, including the secretary problem under matroid and knapsack constraints, and the edge-weighted online bipartite matching problem with random vertex arrivals. For each of these problems, a stream of items arrives in a uniform random order, and items have values that are revealed only when they arrive. The goal is to select a set of items with nearly maximum total value, subject to the constraints (matroid, knapsack, or matching). In the learning-augmented setting, we are also given predictions of item values ahead of time. The goal is to develop algorithms that are both \emph{consistent}, in that accurate predictions lead to near-optimal selection, and \emph{robust}, in that they obtain strong worst-case bounds even when predictions are inaccurate.

Building on work of \citet{FujiiYoshida2024} on the learning-augmented secretary problem, we study a method that follows the predicted optimum until an item with a large enough prediction error is observed. At that point, it reverts to running a worst-case competitive algorithm. We make a simple but critical modification to this algorithm, which selects the first item with a large prediction error with a carefully chosen probability. With this modification, we prove a general result, showing that, letting \(\alpha\in(0,1]\) be the competitive ratio of the underlying worst-case algorithm, this approach achieves competitive ratio
\begin{align*}
\max\left\{\frac{\alpha}{2(1+\alpha)},\frac{1-\eps}{1+\eps}\right\},
\end{align*}
where $\eps$ is the maximum relative error of any prediction. That is, the algorithm is perfectly consistent, achieving competitive ratio $1$ as $\eps \rightarrow 0$, and robust up to a constant factor as compared to the worst-case baseline. For all three problems we study, $\alpha = 1/e$, and our result is the first to achieve consistency and constant-factor robustness. We hope that our general analysis framework can be applied to other learning-augmented online decision-making problems.

\end{abstract}

\section{Introduction}

Online algorithm design attempts to understand how a decision-maker can process a stream of inputs in an incremental fashion, while still making decisions that are nearly as good as if they knew all inputs ahead of time \cite{BorodinElYaniv1998,Albers2003}. Online decision-making arises, e.g., in hiring, where an organization must interview candidates one-by-one and decide to make job offers before seeing all possible candidates \cite{Dynkin1963,Freeman1983,Ferguson1989}. It similarly arises in scheduling, where decisions must be made about how to serve requests as they come in, without the ability to see future requests \cite{Graham1966}. Other application areas include buying and selling assets with changing prices \cite{ElYanivEtAl2001}, allocating internet advertising space \cite{MehtaEtAl2007}, assigning patients to treatment/control groups in clinical trials \cite{VillarEtAl2015}, and beyond \cite{Albers2003}.

Classic worst-case analysis typically assumes that the decision-maker has no knowledge of inputs that have not been seen yet, and aims to develop algorithms with strong guarantees despite this difficulty. However, in many applications, this assumption is too pessimistic: the decision-maker often does have some information about future inputs. E.g., in applications to hiring, they may have estimates of candidate abilities from resumes, portfolios, or references, before conducting interviews. In applications to scheduling, load balancing, or pricing, they may be able to predict statistics about future input sequences using past observations.

Significant work attempts to understand how this noisy information about future inputs can be leveraged to improve outcomes. For example, work on \emph{prophet inequalities} considers the setting where future inputs are drawn independently from known distributions \cite{KrengelSucheston1977,SamuelCahn1984}; see \cite{HillKertz1992,Lucier2017,CorreaEtAl2019} for surveys. Work on \emph{learning-augmented} algorithms takes a more worst-case view. The goal is to design algorithms that are both (1) \emph{consistent}, in that they achieve near-optimal outcomes when predictions about future inputs are accurate and (2) \emph{robust}, in that they achieve non-trivial guarantees even when the information about future inputs is arbitrarily inaccurate. In recent years, significant work has studied learning-augmented algorithms for online decision-making problems, including caching \cite{LykourisVassilvitskii2018}, scheduling \cite{PurohitSvitkinaKumar2018}, metrical task systems \cite{AntoniadisEtAl2020}, purchasing problems like one-way trading \cite{SunEtAl2021}, and selection problems, like the secretary problem \cite{AntoniadisEtAl2023,DuttingEtAl2021,FujiiYoshida2024}.

\subsection{Problem Setup and Existing Results}

In this work, we study several closely related learning-augmented online selection problems, including the secretary problem under general matroid and knapsack constraints and the edge-weighted online bipartite matching problem with random vertex arrivals. For simplicity of exposition, in the introduction, we primarily focus on the secretary problem under general matroid constraints.
\begin{problem}[Learning-Augmented Matroid Secretary]\label{prob:matroid}
The learning-augmented matroid secretary problem is defined as follows:
\begin{itemize}
\item There are $n$ items indexed by the integers $[n]=\{1,\ldots,n\}$, with unknown positive values $v_1,\ldots,v_n$. Each item corresponds to an element of a known matroid $M$.
\item The goal is to select a set $S \subseteq [n]$ that maximizes the total value $ \sum_{i \in S}v_i$ subject to $S$ being an independent set of $M$.
\item The items are assigned uniform and independent random arrival times \(t_1,\ldots,t_n\sim\operatorname{Unif}[0,1]\), and are processed in increasing order of arrival time.\footnote{Sometimes it is assumed that items arrive in a uniform random order, without specified arrival times. This is equivalent---the decision-maker can generate such times when item $i$ arrives.}
\item When an item $i$ arrives, the decision-maker observes its value $v_i$, and must irrevocably decide to add $i$ to the output set (assuming doing so is feasible) or reject it.
\item The decision-maker is given positive predicted values $\vhat_1,\ldots, \vhat_n$ for each item ahead of time. There are no guarantees on the accuracies or distributions of these predictions.
\end{itemize}
\end{problem}
The matroid secretary problem has been heavily studied in the online algorithms literature \cite{BIK2007,BabaioffEtAl2018,OveisGharanVondrak2013,Soto2011,JailletEtAl2013,DinitzKortsarz2013,Lachish2014,FeldmanEtAl2015,BahraniEtAl2021}; see \cite{Dinitz2013} for a survey. If the independent sets of $M$ are exactly the sets of size at most one, we have the classic secretary problem \cite{Ferguson1989}. If they are exactly the sets of size at most $k$, we have the $k$-secretary problem \cite{Kleinberg2005}. Other matroids arise in many natural applications of online selection. E.g., hiring candidates subject to quotas for a set of job categories corresponds to a \emph{partition matroid} constraint; allocating goods to bidders with different preference sets leads to \emph{transversal matroid} constraints; assigning passengers to airline routes subject to capacity constraints leads to \emph{gammoid} constraints, and beyond~\cite{BIK2007,BabaioffEtAl2008}.

In the worst-case setting, the matroid secretary conjecture of \citet{BIK2007}, that a constant-competitive algorithm exists for general matroids, was recently resolved by \citet{Singla2026}, and subsequently the optimal competitive ratio of $1/e$ has been obtained \citep{Huang2026,AbdiEtAl2026}.

A related natural variant of the secretary problem is knapsack secretary~\cite{BabaioffEtAlKnapsack2007}, where each item has a known weight $w_i$ and the selected set $S$ must satisfy $\sum_{i \in S} w_i \le 1$. See \Cref{prob:knapsack}. This models settings where a decision-maker with a fixed budget must choose among options with different costs, such as hiring workers, procuring crowdsourcing services, or bidding for sponsored search ad slots~\cite{BabaioffEtAlKnapsack2007,BadanidiyuruEtAl2012,ZhouEtAl2008}. We will consider its fractional version~\cite{GilibertiKarrenbauer2021,KlimmKnaack2026}, where the decision-maker chooses a fractional allocation $x_i \in [0,1]$ for each item, attempting to maximize $\sum_{i \in [n]} x_i v_i$ subject to $\sum_{i \in [n]} x_i w_i \le 1$.

Finally, we consider an online matching variant in which each element is an edge of a bipartite graph. Vertices on one side of the partition are known in advance, while vertices on the other arrive in a uniform random order. When a vertex arrives, the decision-maker learns the values of all its incident edges, and must irrevocably decide whether to add one of them to the selected set $S$, subject to $S$ being a matching~\cite{KRTV2013}. See \Cref{prob:matching}. Online matching is widely applied, e.g., in online advertising, where the offline vertices are advertisers, the arriving vertices are search queries or ad impressions, and edge values capture the value of showing a given ad in response to a query~\cite{MehtaEtAl2007,FeldmanEtAl2009,Mehta2013}. The problem can also model, e.g., assigning arriving workers to tasks in crowdsourcing markets~\cite{HoVaughan2012} and arriving ride requests to drivers in ride-sharing platforms~\cite{DickersonEtAl2018}.

\medskip

\noindent\textbf{Existing Results.} For all the above problems, in the prediction-free setting, worst-case algorithms (some very recently developed) are known that achieve competitive ratio $1/e$---i.e., the expected total value output by the algorithm is at least a $1/e$ fraction of the offline optimal solution~\cite{AbdiEtAl2026,KlimmKnaack2026,KRTV2013}. Further, for all problems, this bound is known to be tight. However, our understanding in the learning-augmented setting is much more limited.

\citet{FujiiYoshida2024} consider the special cases of \Cref{prob:matroid} corresponding to the classic secretary and $k$-secretary variants. For the classic secretary problem, they give an algorithm with competitive ratio $\max \left (0.215, \frac{1-\eps}{1+\eps} \right ),$ where \vspace{-1em}
\begin{align}\label{eq:eps}
\eps = \max_{i \in [n]} \left | 1 - \frac{\vhat_i}{v_i} \right |
\end{align}
is the maximum relative error of any value prediction. That is, as the predictions become more accurate (i.e., $\eps \rightarrow 0$), the algorithm achieves competitive ratio approaching $1$. Thus, the algorithm is \emph{consistent}. However, even when the error $\eps$ is unbounded, the algorithm achieves a worst-case competitive ratio of $0.215$. Thus, the algorithm is \emph{robust}. Fujii and Yoshida's bound was later slightly improved to $\max \left (0.221,\frac{1-\eps}{1+\eps}\right )$ by \citet{KarisaniEtAl2026}. No algorithm can achieve better than $\max \left (\frac{1}{e},\frac{1-\eps}{1+\eps} \right )$, as $1/e$ is the optimal competitive ratio even when the predictions are completely uninformative. In fact, \citet{ChooLing2024} show a stronger upper bound of $\max \left (\frac{1}{3},\frac{1-\eps}{1+\eps} \right )$---i.e., achieving consistency necessitates a loss in the worst-case competitive ratio. Closing the gap between this upper bound and the existing lower bounds is an interesting open question.

Fujii and Yoshida's algorithm follows the predictions until an item with a large prediction error arrives, and then switches to Dynkin's classical rule~\cite{Dynkin1963}, which also processes the item that triggered the switch. Since this item's position is not uniformly random among the remaining items, the analysis proceeds by a case analysis over the status of the best and best-predicted items. \citet{KarisaniEtAl2026} keep this structure, randomize the decision on the best-predicted item after the switch, and refine the case analysis. Such item-by-item analyses do not directly extend to problems where many items are selected.

\subsection{Our Results}

We significantly generalize the prior work of \citet{FujiiYoshida2024} and \citet{KarisaniEtAl2026}, giving the first consistent and robust algorithms for the learning-augmented secretary problem under general matroid and knapsack constraints, and the edge-weighted online bipartite matching problem. In particular, we show:
\begin{theorem}[Main Result]\label{thm:intro}
For learning-augmented matroid secretary, fractional knapsack secretary, and edge-weighted bipartite matching with random vertex arrivals (\Cref{prob:matroid,prob:knapsack,prob:matching}), there are algorithms that achieve competitive ratio
$$\max \left (\frac{1}{2(e+1)}, \frac{1-\eps}{1+\eps} \right ),$$
where $\eps$ is the maximum relative error among all item or edge predictions.
\end{theorem}
As discussed, for all three problems, it is known that competitive ratio $1/e$ is optimal when no value predictions are given. Thus, our robustness bound of $1/[2(e+1)]$ is a constant factor off the optimum. Whether it can be improved for any or all of the studied problems is an interesting open question. Surprisingly, while \Cref{thm:intro} applies to the fractional knapsack secretary problem, we show in \Cref{sec:integral} that the integral version does not admit a nontrivial robust and consistent algorithm. In particular, any algorithm achieving competitive ratio at least $1-\delta$ when the predictions exactly equal the true values has worst-case competitive ratio at most $\delta$. An algorithm matching this bound up to a constant factor can be obtained simply by running a worst-case algorithm with probability $\delta$ and following the predicted optimum otherwise.

\subsection{Technical Overview}\label{sec:tech-overview}

We prove \Cref{thm:intro} using a simple and generic framework, detailed in \Cref{sec:framework}. Our algorithm (\Cref{alg:general}) follows the approach of \citet{FujiiYoshida2024} for the learning-augmented classic secretary problem. The decision-maker initially `trusts' the predictions: letting $\whx$ be the optimal solution under the predicted values, it selects items according to $\whx$. However, if a sufficiently large prediction error (a `violation') is observed for any arriving item, the decision-maker stops following the predictions, and switches to running a standard worst-case competitive algorithm, denoted $\Base$, on the remainder of the problem instance.

If there are no violating items, it is easy to bound the competitive ratio of the algorithm---all predictions are accurate up to the violation threshold, and thus the predicted optimal solution $\whx$ is near-optimal under the true item values (see \Cref{lem:predicted-optimum-stability} for a formalization). The main analysis thus focuses on the case when there is at least one bad prediction---i.e., at least one violating item. A key difficulty is the first violating item, which triggers the switch to the worst-case algorithm. This item may have high value (e.g., in the classic secretary problem, it may be the best item), so the decision on it may be critical. However, if it is handed to $\Base$, as in \citet{FujiiYoshida2024}, then $\Base$ no longer sees the remaining items in uniformly random order. E.g., if an item is the only violating item, it is always the first arrival handled by $\Base$.

\medskip

\noindent\textbf{A Black-Box Reduction to the Worst-Case Algorithm.} We make a small modification that significantly simplifies Fujii and Yoshida's analysis: we select the first violating item with some fixed probability, and then handle only the later arriving items with $\Base$. Conditioned on all arrivals up to the first violation, these items arrive in a uniform random order, so we can apply the worst-case guarantees of $\Base$ black-box.

Let $f \in [n]$ be the index of the first violating item. Let $\OPT(R)$ be the optimal `residual' value obtainable from all items starting with the first violation (i.e., the maximum value we can obtain by selecting items from the first violation onwards while respecting the feasibility constraints). Let $\OPT(R \setminus f)$ be the optimal residual value for all items \emph{after} the first violation $f$, assuming that we reject $f$. Let $\alpha \in (0,1]$ be the worst-case competitive ratio of $\Base$. Our algorithm will accept $f$ with probability $\frac{\alpha}{\alpha+1}$ (assuming that doing so is feasible), and otherwise reject it. Let $\Delta_f = v_f$ if accepting $f$ is feasible, and $\Delta_f = 0$ otherwise. The accepting branch contributes at least $\Delta_f \cdot \frac{\alpha}{1+\alpha}$ to the expected output value. The rejecting branch contributes at least $\left (1-\frac{\alpha}{\alpha+1}\right ) \cdot \alpha \cdot \OPT(R \setminus f) = \frac{\alpha}{\alpha+1} \cdot \OPT(R\setminus f)$. In total, the expected value of the output is thus at least
\begin{align}\label{eq:techOverview}
 \frac{\alpha}{1+\alpha}\cdot (\Delta_f + \OPT(R \setminus f)) \ge \frac{\alpha}{\alpha+1} \cdot \OPT(R),
\end{align}
where the inequality follows since $\OPT(R)$ obtains at most $\Delta_f$ value from $f$, and its remaining choices are feasible for $R \setminus f$. For all three problems we consider (matroid secretary, knapsack secretary, and online matching), $\alpha = 1/e$. Thus, if we can show that $\E[\OPT(R)] \ge c \cdot \OPT$ for some constant $c \in [0,1]$, we will have a worst-case competitive ratio $\frac{c}{e+1}$ when there is at least one violating item. Combining this with the case of no violating items (\Cref{lem:predicted-optimum-stability}) gives a bound of $\max \left (\frac{c}{e+1}, \frac{1-\eps}{1+\eps} \right )$, as in \Cref{thm:intro}.

\medskip

\noindent\textbf{Residual Value Analysis.} With the above argument in place, the key goal becomes to lower bound the expected residual value $\E[\OPT(R)]$. The proof of this lower bound depends on the specific problem being studied. Here we provide short intuitive arguments that give looser bounds than the final guarantees stated in \Cref{thm:intro}. We later briefly sketch how to obtain the stronger bounds.

\smallskip

\noindent \textit{Online Matching.} Online matching admits a particularly simple argument. Consider any edge $e$ in the true optimal matching whose offline endpoint is matched by some edge $\wh e$ in the predicted optimal matching. Note that $e$ is available and feasible to include in $\OPT(R)$ exactly when the online endpoints of both $e$ and $\wh e$ arrive at or after the first violation $f$. Further, this event happens with probability at least $1/3$: the worst case is when $\wh e$ and $e$ are both non-violating, and there is a single violating item $f$. The event happens exactly when $f$ arrives first out of the three items---i.e., with probability $1/3$. If the online endpoint of $e$ or $\wh e$ is violating, it must arrive at or after $f$ (since $f$ is the first violation) and so the probability only increases. Thus, in expectation, edges carrying at least a $1/3$ fraction of the optimal value are feasible to include in $\OPT(R)$, and so we have $\E[\OPT(R)] \ge 1/3 \cdot \OPT$. Combined with \eqref{eq:techOverview} this yields a $\max \left (\frac{1}{3(e+1)}, \frac{1-\eps}{1+\eps} \right )$ competitive ratio bound for online matching.

\smallskip

\noindent\textit{Matroid Secretary.} The analysis for matroid secretary is a bit more subtle---unlike in matching, we cannot consider elements individually, as groups of elements in the optimum solution may be jointly blocked by predicted-optimum elements, even if they are not individually blocked.

Let $H$ be the set of `violating' items with large prediction error, let $\Ohat$ be the predicted optimal basis, and let $O$ be the true optimal basis. Let $T$ denote the arrival time of the first violation, and condition on $T = t$. By then, the algorithm has accepted a set $A\subseteq\Ohat$ and discarded a set $D$ of items outside $\Ohat$. Fix any $i$, and let $O_i$ be the $i$ highest-value items of $O$. Let $U=O_i\setminus D$ be the subset of these items that are not discarded before the first violation. Note that $\E[|U| \mid T = t] \ge (1-t) |O_i|$, since each item arrives before time $t$ with probability at most $t$ (probability exactly $t$ for non-violating items, and probability $0$ for violating items, since $t$ is the time of the first violation). While items in $U$ have not been discarded before the first violation, they may be blocked by the accepted items $A$. We next bound the probability of this occurring. By the basis exchange property (see \Cref{sec:preliminaries}), there is a basis $B$ with $U\subseteq B\subseteq U\cup\Ohat$. Let $C=\Ohat\setminus B$ be the set of `conflicting' items in $\Ohat$ that would potentially block items in $U$. Accepted items in $\Ohat$ but outside $C$ lie in $B$, so together with $U$ they form an independent set and block nothing in $U$. On the other hand, each accepted item in $C$ blocks at most one item of $U$. In addition, an item of $U\cap\Ohat$ that arrives before the first violation is itself accepted, so it is no longer available in the residual. We can observe that $|C|=|U\setminus\Ohat|$: to form the basis $B$ with $U \subseteq B$ we remove exactly the items $C$ from $\Ohat$, replacing them with the items $U \setminus \Ohat$. Thus, we have $|C|+|U\cap\Ohat|=|U|$. Each item of $C$ and each item of $U\cap\Ohat$ arrives before the first violation, and is thus accepted, with probability at most $t$. So the expected number of items of $U$ that are blocked or already accepted is at most $t|U|$. Overall, the expected number of items from $O_i$ available for use in $\OPT(R)$, conditioning on a first violation at time $T = t$, is at least $(1-t) \E[|U| \mid T = t] \ge (1-t)^2 |O_i|$. Applying this bound to $O_i$ for every $i$, and using that $\E[(1-T)^2]\ge 1/3$, gives $\E[\OPT(R)] \ge \OPT/3$.

\smallskip

\noindent\textit{Knapsack Secretary.} Finally, for fractional knapsack secretary, we let $W$ be the capacity consumed by following the predicted optimal solution before the first violation. We let $L$ be the total value that the optimal allocation obtains from items that arrive before the first violation. If we restrict the optimal allocation just to items arriving after the first violation, it has value $\OPT-L$ and uses capacity at most $1$. Scaling this allocation by the remaining capacity $1-W$ therefore gives a feasible residual allocation, so $\OPT(R)\ge(1-W)(\OPT-L)$.

Conditioning on the first violation arriving at time $T=t$, each non-violating item arrives before $t$ with probability $t$, while no violating item does. Thus, $\E[W\mid T=t]\le t$ and $\E[L\mid T=t]\le t\OPT$. Moreover, conditional on $T=t$, $W$ and $L$ are positively correlated: both are nonnegative weighted sums of the same independent indicators for whether an item arrives before $t$. It follows that
\[
\E[\OPT(R)\mid T=t]
\ge (1-\E[W\mid T=t])(\OPT-\E[L\mid T=t])
\ge (1-t)^2\OPT.
\]
Again using that $\E[(1-T)^2]\ge1/3$ gives $\E[\OPT(R)]\ge\OPT/3$.

\medskip

\noindent\textbf{Tightness of Residual Bound.} Unfortunately, it is not hard to see that the bound of $\E[\OPT(R)]\ge\OPT/3$ cannot be significantly improved. All three problems studied capture the standard secretary problem as special cases: matroid secretary when only sets of size at most one are feasible, knapsack secretary when all items have weight one and so exactly one can be accepted into the knapsack, and bipartite matching when there is just a single fixed vertex that all arriving edges share as an endpoint. For the classic secretary problem, analogously to the worst case in the online matching argument above, if we have just a single low-value violating item $f$, along with optimal and predicted optimal items $o$ and $\hat o$, $o$ is only available to include in $\OPT(R)$ if $f$ arrives first---which happens with probability $1/3$. If $f$ arrives second, either $o$ is rejected before the first violation, or $\hat o$ is accepted, meaning that $o$ cannot be later accepted. If $f$ arrives last, then there are no remaining items in the residual. Given this, we have $\E[\OPT(R)] \approx \OPT/3$. A slightly better bound can indeed be proven since if neither $o$ nor $\hat o$ is a violating item, they must have relatively comparable values. However, it is not strong enough to give the $\max \left (\frac{1}{2(e+1)}, \frac{1-\eps}{1+\eps} \right )$ competitive ratio of \Cref{thm:intro}.

\medskip

\noindent\textbf{Tighter Bounds via Prefix + Residual Analysis.} Given this, the analysis of \Cref{thm:intro} must look beyond just $\OPT(R)$. In particular, we need to use that when $\OPT(R)$ is small, it is because the algorithm has already collected value before the first violation, and this value must be accounted for. We let $P$ denote the `prefix' value obtained from all items in the predicted optimum accepted before the first violation. As discussed, our algorithm obtains expected value at least $\frac{1}{1+e} \OPT(R)$ on all items arriving at or after the first violation. Thus, its expected total value is at least $\E[P + \frac{1}{1+e} \OPT(R)]$. See \Cref{lem:first-violation-decision} for a formalization. In \Cref{sec:applications} we show that for all three problems, this total expected prefix+residual value can be lower bounded by $\frac{1}{2(e+1)} \cdot \OPT$, yielding the bound of \Cref{thm:intro}. Very roughly, in all three cases, we argue that $\OPT(R)$ being small makes $P$ large, obtaining a stronger bound than when bounding $\E[\OPT(R)]$ alone.

\section{Preliminaries}\label{sec:preliminaries}

We start with preliminaries relevant to our results on all three problems covered by \Cref{thm:intro}.

\subsection{Problem Definitions}

As discussed, we study the learning-augmented matroid secretary problem (\Cref{prob:matroid}), as well as the closely related fractional knapsack secretary problem and edge-weighted online bipartite matching problem with random vertex arrivals. We define the latter two problems below.

\begin{problem}[Learning-Augmented Fractional Knapsack Secretary]\label{prob:knapsack}
The learning-augmented fractional knapsack secretary problem is identical to the learning-augmented matroid secretary problem (\Cref{prob:matroid}) with the following differences:
\begin{itemize}
\item Each item has a weight $w_i\in(0,1]$ known ahead of time to the decision-maker.
\item The goal is to select fractions $x_i\in[0,1]$ that maximize the total value $\sum_{i\in[n]}x_iv_i$ subject to the knapsack constraint $\sum_{i\in[n]}x_iw_i\le1$.
\item When an item arrives, the decision-maker observes its value $v_i$ and must irrevocably choose its fraction $x_i$, subject to the capacity left by earlier decisions.
\end{itemize}
\end{problem}
The integral version, which we discuss in \Cref{sec:integral}, further requires $x_i\in\{0,1\}$ for every item.

\begin{problem}[Learning-Augmented Online Bipartite Matching with Random Arrivals]\label{prob:matching}
The learning-augmented online bipartite matching problem with random arrivals is defined as follows:
\begin{itemize}
\item There is a known bipartite graph $G=([n]\cup U,E)$, with unknown positive edge values $v_e$ for $e\in E$.
\item The goal is to select a set $S\subseteq E$ that maximizes the total value $\sum_{e\in S}v_e$ subject to $S$ being a matching, i.e., no two selected edges share an endpoint.
\item The vertices in $[n]$ are assigned uniform and independent random arrival times \(t_1,\ldots,t_n\sim\operatorname{Unif}[0,1]\), and processed in increasing order of arrival time. When a vertex arrives, the decision-maker observes the values of all its incident edges, and must irrevocably decide to add at most one of these edges to $S$, subject to the matching constraint, or reject the vertex.
\item The decision-maker is given predicted values $\vhat_e$ for every edge $e \in E$ ahead of time. There are no guarantees on the accuracies or distributions of these predictions.
\end{itemize}
\end{problem}

\subsection{Notation and Prediction Error}

For matroid and knapsack, items are indexed by $[n]$; for matching, arriving vertices are indexed by $[n]$ but edges are indexed by $E$. Let $\mathcal C$ denote the corresponding coordinate set: $\mathcal C=[n]$ for matroid and knapsack and $\mathcal C=E$ for matching. For all three problems, arrivals are indexed by $[n]$. We let $V_i$ denote the set of coordinates whose values are revealed at arrival $i$. In matroid and knapsack secretary, $V_i$ is simply the singleton containing element $i$. For online matching, $V_i$ is the set of edges incident to the arriving vertex $i$.

We write $\mathbf v$ and $\widehat{\mathbf v}$ for the true and predicted values on the coordinates in $\mathcal C$. Each problem requires selecting a decision vector $\bv x$, with $x_c\in\{0,1\}$ for matroid elements and matching edges and $x_c\in[0,1]$ for fractional-knapsack items. Let $\mathcal X$ be the set of decision vectors satisfying the corresponding feasibility constraints. In all cases, $\mathcal X$ is known to the decision-maker in advance.

A decision vector $\mathbf x\in\mathcal X$ has value $\mathbf v\cdot\mathbf x$. We denote the predicted and true optimum decisions by
\begin{align}
\widehat{\mathbf x} \in\arg\max_{\mathbf x\in\mathcal X} \widehat{\mathbf
v}\cdot\mathbf x, \qquad \mathbf x^* \in\arg\max_{\mathbf x\in\mathcal X}
\mathbf v\cdot\mathbf x, \qquad \OPT=\mathbf v\cdot\mathbf x^*.
\end{align}
We assume that a deterministic tie-breaking rule guarantees a unique optimum for any value vector.

The maximum relative prediction error $\eps$ is defined in \eqref{eq:eps}, with the maximum taken over all coordinates in $\mathcal C$. We work with the equivalent definition: $\eps$ is the minimum value such that
\begin{equation}\label{eq:prediction-error}
 (1-\eps)v_c\le\vhat_c\le(1+\eps)v_c
\qquad\text{for every }c\in\mathcal C.
\end{equation}
If the prediction error $\eps$ is small, then following the predicted optimum gives a near-optimal solution. In particular, we can observe:
\begin{lemma}[Value of the Predicted Optimum]\label{lem:predicted-optimum-stability} For maximum prediction error $\eps<1$, the predicted optimal solution $\widehat{\mathbf x}$ has value at least $\bv v \cdot \widehat{\mathbf x} \ge \frac{1-\eps}{1+\eps} \cdot \OPT$.
\end{lemma}
\begin{proof}
Using \eqref{eq:prediction-error} we can bound:
\[
 \mathbf v\cdot\widehat{\mathbf x}
 \ge \frac{1}{1+\eps}\widehat{\mathbf v}\cdot\widehat{\mathbf x}
 \ge \frac{1}{1+\eps}\widehat{\mathbf v}\cdot\mathbf x^*
 \ge \frac{1-\eps}{1+\eps}\mathbf v\cdot\mathbf x^*
 = \frac{1-\eps}{1+\eps}\OPT.
\]
\end{proof}

As discussed, our general algorithm will follow the predicted optimum $\whx$ until it sees an item with a sufficiently large prediction error. We call such an item a `violation'.
\begin{definition}[Violation]\label{def:violation}
Fix a threshold $\theta\in[0,1)$. In matroid and knapsack secretary, item $i\in[n]$ is a violation if $\vhat_i\notin[(1-\theta)v_i,(1+\theta)v_i]$. In online matching, an arriving vertex is a violation if at least one of its incident edges fails the same inequality. Let $H$ denote the set of violating items or arriving vertices.
\end{definition}
In \Cref{alg:general}, we will fix the violation threshold to $\theta=\frac{2e+1}{2e+3}$, which is optimized for the competitive ratio of $1/e$ achieved by the worst-case algorithm for all three of the problems we study. In particular, for this setting of $\theta$, $\frac{1-\theta}{1+\theta} = \frac{1}{2(e+1)}$, which is the final worst-case competitive ratio of our algorithm.

\paragraph{Additional Notation and Terminology.} For the matroid secretary and matching problems we sometimes work directly with a set of matroid elements or edges $S$, as opposed to its corresponding decision vector. We denote $v(S)=\sum_{i\in S}v_i$ and $\vhat(S)=\sum_{i\in S}\vhat_i$. We say that an algorithm has competitive ratio $\rho$ if $\E[\ALG]\ge\rho\OPT$ on every instance, where $\E[\ALG]$ is the expected value of the algorithm output, taken over the random arrival sequence and any randomness of the algorithm.

\paragraph{Matroid Preliminaries.} Write $M=([n],\I)$ for the known matroid in \Cref{prob:matroid}, where $\I$ is its family of independent sets. A basis is a maximal independent set. All bases have the same size, and every independent set can be extended to a basis. Since values are positive, both the predicted and true optima are bases.

If an independent set $A$ has already been accepted, its contraction $M/A$ has independent sets $J$ such that $J\cap A=\varnothing$ and $A\cup J\in\I$. We let $M \setminus D$ denote the operation of removing a set $D$ of elements from consideration. All elements in $D$ are removed from the ground set, and any independent set containing an element of $D$ is removed from $\mathcal I$. With this notation, $(M/A)\setminus D$ is precisely the remaining matroid after accepting the elements in $A$ and discarding the elements in $D$.

Our analysis uses two standard basis exchange properties for matroids~\cite[Chapter~39]{Schrijver2003}. Given a basis $B$ and an independent set $J$, there is a basis $B'$ with $J\subseteq B'\subseteq J\cup B$. I.e., $B'$ completes $J$ to a basis using only elements from $B$. Also, for any two bases $B$ and $B'$, there is a bijection $\phi:B'\setminus B\to B\setminus B'$ such that $(B\setminus\{\phi(i)\})\cup\{i\}$ is a basis for every $i\in B'\setminus B$.

\section{General Algorithm and Analysis}\label{sec:framework}

Pseudocode for the switching algorithm is given in \Cref{alg:general}.

\begin{algorithm}[h]
\caption{General Switching Algorithm}\label{alg:general}
\begin{algorithmic}[1]
\Require Feasible set $\mathcal X$, predictions $\hv$, and a baseline algorithm $\Base$.
\State Compute $\whx\in\arg\max_{\bv y\in\mathcal X}\hv\cdot\bv y$. Set $\bv x\gets\bv 0$ and $\mathit{mode}\gets\mathsf{PRED}$.
\While{$\mathit{mode}=\mathsf{PRED}$}
    \State Observe the next arrival $j$. Let $V_j$ be the set of coordinates whose values are revealed at $j$.
    \If{$j$ is not a violation with threshold $\theta = \frac{2e+1}{2e+3}$ (\Cref{def:violation})}
        \State Set the coordinates of $\bv x$ in $V_j$ according to $\whx$.
    \Else
        \State With probability $p$, take the feasible action on $V_j$ maximizing the value added; otherwise reject $j$.
        \State Initialize $\Base$ on later arrivals under the remaining constraints. Set $\mathit{mode}\gets\mathsf{BASE}$.
    \EndIf
\EndWhile
\While{$\mathit{mode}=\mathsf{BASE}$}
    \State Process the next arrival using $\Base$.
\EndWhile
\State \Return $\bv x$.
\end{algorithmic}
\end{algorithm}

As discussed, \Cref{alg:general} starts in $\mathsf{PRED}$ mode, following the predicted optimum $\whx$, until the first violating arrival, i.e., the first arrival with relative prediction error larger than $\theta = \frac{2e+1}{2e+3}$. The algorithm handles this first violating arrival with a random choice: with probability $p$ (eventually set to $\frac{1}{1+e}$ for the three problems studied), it takes the feasible action on the arrival that maximizes total value. Otherwise, it rejects the arrival. For matroid secretary, this means it accepts the violating item $j$ with probability $p$, and rejects it otherwise. For fractional knapsack, with probability $p$ it sets $x_j$ to the largest allocation that does not violate feasibility, i.e., $x_j=\min\{1,C/w_j\}$, where $C$ is the remaining capacity. Otherwise, it rejects the item, setting $x_j = 0$. For online matching, with probability $p$, it selects the maximum-value edge from vertex $j$ to an unmatched offline vertex (assuming one exists) and otherwise rejects the vertex.

After handling the first violation, the algorithm enters $\mathsf{BASE}$ mode, where it simply runs a worst-case algorithm \(\Base\) on later arrivals, subject to the modified constraints given the decisions before the first violation. In all cases, the modified problem is just an instance of the original. For matroid secretary, it is a matroid secretary problem on a contracted matroid. For knapsack secretary, it is a knapsack secretary problem with a smaller capacity.\footnote{To handle a smaller capacity $C$ with a standard algorithm, we simply divide all weights and the capacity by $C$ to normalize the capacity to $1$ without changing the value of any feasible allocation. If $C=0$, we simply reject all remaining items.} For online matching, it is an online matching problem on a subset of the vertices (those that have not been previously matched). Further, conditioned on the arrivals up to the first violation $j$, the order of the later arrivals is still uniform, and their arrival times are uniformly distributed in $[t_j,1]$. Thus, in all instances, $\Base$ with competitive ratio $1/e$ is available~\cite{AbdiEtAl2026,KlimmKnaack2026,KRTV2013}.

\subsection{Analysis}

Observe that if no violation occurs (i.e., all predictions are accurate with threshold $\theta$), then the maximum prediction error satisfies $\eps \le \theta = \frac{2e+1}{2e+3}$ (as set in \Cref{def:violation}). So,
$$\frac{1-\eps}{1+\eps} \ge \frac{1 - \frac{2e+1}{2e+3}}{1+\frac{2e+1}{2e+3}} = \frac{1}{2(e+1)}.$$
In this case, by \Cref{lem:predicted-optimum-stability}, \Cref{alg:general} thus obtains competitive ratio $\frac{1-\eps}{1+\eps} = \max \left (\frac{1}{2(e+1)} ,\frac{1-\eps}{1+\eps}\right ),$ as required by \Cref{thm:intro}.

Thus, the challenging case to analyze is when there is at least one violation and \Cref{alg:general} enters $\mathsf{BASE}$ mode. In this case, we will analyze the total expected value of the \emph{prefix} decided before the first violation and the \emph{residual} decided after the first violation, formally defined below.

\begin{definition}[Prefix and Residual]\label{def:resid}
Consider an input instance to \Cref{alg:general} containing at least one violating prediction (i.e., for which $\mathsf{BASE}$ mode is guaranteed to be triggered). Let $f$ be the index of the first violating arrival and let $\bv x_f$ be the value of $\bv x$ immediately before this arrival is processed. Let $\mathcal{X}' \subseteq \mathcal{X}$ be the set of feasible decision vectors that match $\bv x_f$ on all items revealed before arrival $f$. We define:
$$P = \bv x_f \cdot \bv v\hspace{2em}\text{and}\hspace{2em}\OPT(R) = \max_{\bv x \in \mathcal{X}'} (\bv x-\bv x_f) \cdot \bv v.$$
That is, $P$ is the value obtained by all decisions before arrival $f$. $\OPT(R)$ is the maximum \emph{additional obtainable} value using all items revealed at arrival $f$ or later. Note that both $P$ and $\OPT(R)$ are random variables, depending on the random arrival sequence.
\end{definition}

To show that \Cref{alg:general} satisfies the competitive ratio of \Cref{thm:intro}, we will argue that $P$ and $\OPT(R)$ are sufficiently large in expectation. At first glance, one might hope that the algorithm obtains expected total value $\E[P + 1/e \cdot \OPT(R)]$, since the worst-case algorithm $\Base$ has competitive ratio $1/e$ for all problems studied. However, this is not quite the case: $\OPT(R)$ potentially includes items revealed by the first violating arrival $f$. While the arrival order of all items arriving \emph{after} $f$ is uniform, since $f$ is guaranteed to be the first arriving item in the residual instance, we cannot simply apply $\Base$ to it.

Consider a simple example of the classic secretary problem: assume that the only violating item is the true maximum-value element $i^* = \arg\max_{i \in [n]} v_i$. All other predictions are accurate. Thus, we always have $f = i^*$. If $\Base$ is the classic $1/e$-competitive algorithm for the secretary problem, it would \emph{always} reject $i^*$, since it always rejects the first $1/e$ fraction of elements seen.

\Cref{alg:general} avoids this issue by making a separate random decision for the first violating arrival $f$, before applying $\Base$ to all later arrivals. The random decision ensures the following bound on the algorithm's competitive ratio in terms of the prefix and residual values:

\begin{lemma}[Prefix and Residual Bound]\label{lem:first-violation-decision}
Suppose \Cref{alg:general} is implemented with \(\Base\) that obtains worst-case competitive ratio $1/e$ and $p = 1/(1+e)$. Then on any input instance containing at least one violating prediction,
$$\E[\ALG] \ge \E [P + p \cdot \OPT(R)].$$
\end{lemma}
\begin{proof}
Condition on the history through the arrival of $f$, before its random decision. Let $\Delta_f$ be the maximum additional value obtainable from the indices $V_f$ while keeping the prefix decisions fixed. For matroid secretary, $\Delta_f=v_f$ if adding $f$ preserves independence, and $\Delta_f=0$ otherwise. For fractional knapsack, $\Delta_f=v_f\min\{1,C/w_f\}$, where $C$ is the remaining capacity. For online matching, $\Delta_f$ is the maximum value of an edge from $f$ to an unmatched offline vertex, or $0$ if no such edge exists.

Let $R\setminus f$ be the residual instance after rejecting $f$, with only the prefix decisions fixed. The later arrivals have uniformly random relative order, so $\Base$ obtains expected value at least $\OPT(R\setminus f)/e$ on this instance. The accepting branch contributes at least $\Delta_f$, since its subsequent value is nonnegative. Averaging over the two branches gives
\begin{align}\label{eq:cam1}
\E[\ALG]\ge\E\left[P+p\cdot \Delta_f+\frac{1-p}{e}\cdot\OPT(R\setminus f)\right].
\end{align}
Any solution in $R$ obtains at most $\Delta_f$ from $V_f$, and its remaining choices are feasible in $R\setminus f$. Thus $\OPT(R)\le \Delta_f+\OPT(R\setminus f)$. Using that $p = 1/(1+e)$ and so $(1-p)/e=p$, we conclude
\[
\E[\ALG]\ge\E\left[P+p\bigl(\Delta_f+\OPT(R\setminus f)\bigr)\right]
\ge\E\left[P+p\cdot\OPT(R)\right],
\]
as desired.
\end{proof}

With \Cref{lem:first-violation-decision} in place, to prove \Cref{thm:intro} it suffices to argue that $\E[P + p \cdot \OPT(R)]$, which lower bounds $\E[\ALG]$, is sufficiently large. In particular, we have:

\begin{lemma}[Reduction to the Prefix and Residual]\label{lem:first-error-guarantee}
Suppose \Cref{alg:general} is implemented with \(\Base\) that obtains worst-case competitive ratio $1/e$ and $p = 1/(1+e)$. Assume that on any given problem instance with at least one violating prediction
  \begin{equation}\label{eq:abstract-weighted-recovery}
    \E\left[P+p\cdot \OPT(R)\right]\ge\frac p2 \cdot \OPT.
  \end{equation}
  Then \Cref{alg:general} achieves competitive ratio
  \begin{equation*}
 \max \left (\frac{1}{2(e+1)}, \frac{1-\eps}{1+\eps} \right ).
\end{equation*}
\end{lemma}

\begin{proof}
Fix the input instance. If there is no violating prediction, the claimed competitive ratio was established at the beginning of this subsection, using \Cref{lem:predicted-optimum-stability}.

Otherwise there is at least one violating prediction, i.e., $\eps > \theta = \frac{2e+1}{2e+3}$. Then by \Cref{lem:first-violation-decision} and the assumption of \eqref{eq:abstract-weighted-recovery} we have:
\[
\E[\ALG]\ge \E\left[P+p\OPT(R)\right]\ge\frac p2 \cdot \OPT = \frac{1}{2(e+1)} \cdot \OPT = \max \left (\frac{1}{2(e+1)} ,\frac{1-\eps}{1+\eps}\right ) \cdot \OPT,
\]
as desired. This completes the proof.
\end{proof}

With \Cref{lem:first-error-guarantee} in place, to prove our main result, \Cref{thm:intro}, it suffices to show that \eqref{eq:abstract-weighted-recovery} holds for our three problems of interest. We do this in \Cref{sec:applications}.

\begin{remark}[Tightness of \eqref{eq:abstract-weighted-recovery}]\label{rem:prefix-residual-tight}
We note that the bound of \eqref{eq:abstract-weighted-recovery} cannot be improved for any of the three problems. As in the discussion of the residual bound in \Cref{sec:tech-overview}, it suffices to consider the standard secretary problem, which is a special case of all three. Suppose there is an item $o$ of value $1$ whose prediction is correct, along with a single violating item $f$ that has negligible true value but whose predicted value exceeds $1$. Then the predicted optimum consists of $f$ alone, so the algorithm accepts nothing before the first violation, and $P=0$. Moreover, $o$ is available in the residual instance only if $f$ arrives first, which happens with probability $1/2$; otherwise $o$ has already been rejected. Thus $\E[P+p\cdot\OPT(R)]\approx\frac p2\cdot\OPT$. Improving the robustness guarantee of \Cref{thm:intro} would therefore require going beyond our prefix + residual analysis.
\end{remark}

\section{Applications to the Three Problems}\label{sec:applications}

We now verify \eqref{eq:abstract-weighted-recovery} for each of the three problems. Together with \Cref{lem:first-error-guarantee}, this proves \Cref{thm:intro}.

\subsection{Matroid Secretary}\label{sec:matroid}

We start with the matroid secretary problem (\Cref{prob:matroid}).

\smallskip
\noindent\textbf{Problem-Specific Notation.} Let $M=([n],\I)$ be the known matroid. For a set of items $S$, write $v(S)=\sum_{i\in S}v_i$ and $\vhat(S)=\sum_{i\in S}\vhat_i$ for its true and predicted values. Let $\Ohat$ and $O$ be independent sets of maximum predicted and true value. Since all values are positive, both are bases.

To prove \eqref{eq:abstract-weighted-recovery}, we consider an instance with nonempty violating item set $H$ (\Cref{def:violation}). We write $h=|H|$. Let $f$ be the first violating item to arrive and let $T$ be its arrival time; both are random. Every item arriving before $T$ is non-violating, so before $T$ the algorithm accepts exactly the items of $\Ohat$ that arrive. The accepted set is thus $A=\{\hat o\in\Ohat:t_{\hat o}<T\}$ and the discarded set is $D=\{i\in[n]\setminus\Ohat:t_i<T\}$. The residual matroid is $R=(M/A)\setminus D$, and it contains $f$. If $\{f\}$ is not independent in $R$, \Cref{alg:general} rejects $f$; otherwise it accepts $f$ with probability $p$. In this notation, the prefix and residual values of \Cref{def:resid} are
\begin{align}\label{eq:matroidPAndR}
P = v(A) \hspace{2em}\text{and}\hspace{2em}\OPT(R) = \max_{J \text{ independent in }R} v(J).
\end{align}

Let $N\defeq[n]\setminus(\Ohat\cup H)$ be the non-violating items outside $\Ohat$. Since violating items arrive at or after $T$, we always have $A\subseteq\Ohat\setminus H$ and $D\subseteq N$. Moreover, conditioned on $T=t$, each item of $\Ohat\setminus H$ belongs to $A$ with probability $t$, and each item of $N$ belongs to $D$ with probability $t$. This holds because $T$ is determined by the arrival times of the items in $H$, which are independent of the arrival times of all other items. Since $A$ and $D$ are determined by the arrival times of disjoint sets of items, they are also independent given $T=t$.

The main result of this subsection is that \eqref{eq:abstract-weighted-recovery} holds for the matroid secretary problem.

\begin{theorem}\label{thm:matroid-weighted-recovery}
For any input instance to \Cref{prob:matroid} with at least one violating item (i.e., with \(H\ne\varnothing\)), letting $P$ and $\OPT(R)$ be as in \eqref{eq:matroidPAndR},
\begin{equation}\label{eq:matroid-weighted-recovery}
  \E\bigl[P+p\OPT(R)\bigr]
  \ge\frac p2\OPT.
\end{equation}
\end{theorem}

To prove \Cref{thm:matroid-weighted-recovery}, we want to show that every element of the true optimum $O$ receives expected weight at least $p/2$ in $\E[P+p\OPT(R)]$. We do this in two steps. First, we drop the factor $p$ and lower bound $\E[P+\OPT(R)]$. \Cref{lem:matroid-layers,lem:matroid-prefix} give a separate guarantee for each element $o\in O$: in expectation, $o$ is either accepted before the first violation or still available to the residual solution, with a probability that depends only on whether $o$ is in $\Ohat$ and whether it is violating.

Second, we bring back the factor $p$ by writing $P+p\OPT(R)=p\,(P+\OPT(R))+(1-p)P$. The first term is handled by step one, and the second adds extra weight on items accepted before the first violation. Combined, we show that this gives every element of $O$ weight at least $p/2$, except in one case: when there is exactly one violating item, an element of $O\setminus\Ohat$ that is not violating only gets weight $p/3$. To cover this shortfall, we pair each such element of $O$ with its blocking element in $\Ohat$. Since both items are non-violating, we can argue that they have comparable values, and so we can use the probability of accepting the blocking item in the prefix to make up for the shortfall in the probability of accepting the true optimal item.

\begin{lemma}\label{lem:matroid-layers}
Consider the setting of \Cref{thm:matroid-weighted-recovery}. Let $B$ be a basis of $M$ with $\Ohat\setminus H\subseteq B$, and for $o\in O$ let $a_B(o)=\one_{\{o\notin B\}}+\one_{\{o\in N\}}$. Then for every $t\in[0,1]$,
\[
\E\bigl[v(A\cap O)+\OPT(R)\mid T=t\bigr]\ge\sum_{o\in O}(1-t)^{a_B(o)}\,v_o.
\]
\end{lemma}
In our main application of \Cref{lem:matroid-layers}, we will take $B = \Ohat$. In this case, $a_{\Ohat}(o) \in \{0,1,2\}$ is exactly the number of ways in which $o$ may become unavailable: by being blocked by an element of $\Ohat$ accepted before the first violation, or by being discarded before the first violation.

\begin{proof}
Order $O=\{o_1,\ldots,o_r\}$ so that $v_{o_1}\ge\cdots\ge v_{o_r}$, write $O_i=\{o_1,\ldots,o_i\}$, and set $v_{o_{r+1}}=0$ and $\Delta_i=v_{o_i}-v_{o_{i+1}}\ge0$. Let $\rk_R(Q)$ denote the maximum size of an independent subset of $Q$ in $R$, and define
\[
q_i\defeq|O_i\cap A|+\rk_R\bigl(O_i\setminus(A\cup D)\bigr).
\]
That is, $q_i$ is the number of elements of $O_i$ accepted before the first violation, plus the number of further elements of $O_i$ that can still be added in $R$. We first show that for every $i$,
\begin{equation}\label{eq:matroid-rank-layer}
\E[q_i\mid T=t]\ge\sum_{o\in O_i}(1-t)^{a_B(o)}.
\end{equation}

Fix $i$ and condition on $T=t$ and on $D$. Let $U=O_i\setminus D$. Since $U\subseteq O$ is independent, there is a basis $B'$ with $U\subseteq B'\subseteq U\cup B$. Let $C=B\setminus B'$ be the set of `conflicting' items: those of $B$ that must be removed to complete $U$ to a basis, and that may therefore block items of $U$. Both $B'$ and $C$ are determined by $D$ and $i$, so they are fixed once we condition on $D$. Because $A\subseteq B$, we have $A\setminus C=A\cap B'$, so $U\cup(A\setminus C)\subseteq B'$ is independent. Because $A\cap D=\varnothing$, we have $O_i\cap A=U\cap A$ and $O_i\setminus(A\cup D)=U\setminus A$. Deleting $D$ does not change the rank of a set disjoint from $D$, and $A$ is independent, so
\[
q_i=|U\cap A|+\rk_{M/A}(U\setminus A)=|U\cap A|+\rk_M(U\cup A)-|A|\ge|U\cap A|+|U\cup(A\setminus C)|-|A|.
\]
Since $U\subseteq B'$ is disjoint from $C$, we have $|U\cup(A\setminus C)|=|U|+|A|-|A\cap C|-|U\cap A|$, and therefore $q_i\ge|U|-|A\cap C|$.

Now average over $A$. Only items of $\Ohat\setminus H$ can belong to $A$, and each does so with probability $t$, so $\E[|A\cap C|\mid D,T=t]\le t|C|$. Since $B$ and $B'$ are bases of equal size and $B'\subseteq U\cup B$, we have $|C|=|B'\setminus B|=|U\setminus B|$. Hence
\[
\E[q_i\mid D,T=t]\ge|U|-t|U\setminus B|=|U\cap B|+(1-t)|U\setminus B|=\sum_{o\in U}(1-t)^{\one_{\{o\notin B\}}}.
\]
Finally, average over $D$. An element $o\in O_i$ belongs to $U$ with probability $1-t$ if $o\in N$, and with probability one otherwise. This gives \eqref{eq:matroid-rank-layer}.

Next we convert these bounds into a bound on value. Let $J$ be the output of the greedy algorithm that scans $o_1,\ldots,o_r$ in this order, skipping elements of $A\cup D$, and adds an element whenever $J$ stays independent in $R$. After the elements of $O_i$ have been scanned, $J\cap O_i$ is a maximal independent subset of $O_i\setminus(A\cup D)$ in $R$, and it is not changed by later steps. In a matroid, all maximal independent subsets of a set have the same size, so $|J\cap O_i|=\rk_R(O_i\setminus(A\cup D))$. For any $Q\subseteq O$ we have $v(Q)=\sum_{i=1}^r\Delta_i|Q\cap O_i|$, because $\sum_{i\ge j}\Delta_i=v_{o_j}$. Applying this to $A\cap O$ and $J\subseteq O$, and using that $J$ is independent in $R$, we get
\[
v(A\cap O)+\OPT(R)\ge v(A\cap O)+v(J)=\sum_{i=1}^r\Delta_i\bigl(|A\cap O_i|+|J\cap O_i|\bigr)=\sum_{i=1}^r\Delta_iq_i.
\]
Taking the conditional expectation, applying \eqref{eq:matroid-rank-layer}, and using $\Delta_i\ge0$, the right side is at least $\sum_{i=1}^r\Delta_i\sum_{o\in O_i}(1-t)^{a_B(o)}=\sum_{o\in O}(1-t)^{a_B(o)}v_o$, where the last step again uses $\sum_{i\ge j}\Delta_i=v_{o_j}$.
\end{proof}

Taking $B=\Ohat$ and averaging over $T$ gives the following bound on the expected prefix plus the full residual value.

\begin{lemma}\label{lem:matroid-prefix}
Consider the setting of \Cref{thm:matroid-weighted-recovery}, and let $h=|H|$ be the number of violating items. Then
\begin{align}
  \E\bigl[P+\OPT(R)\bigr] \ge{}&v(O\cap \Ohat)
  +\frac{h}{h+1} \cdot v((O\setminus \Ohat)\cap H)\notag\\
  &+\frac{h}{h+2}\cdot v((O\setminus \Ohat)\setminus H)
  +\frac1{h+1} \cdot v((\Ohat\setminus O)\setminus H).
  \label{eq:matroid-prefix-bound}
\end{align}
\end{lemma}

\begin{proof}
Apply \Cref{lem:matroid-layers} with $B=\Ohat$. Then $a_{\Ohat}(o)$ equals $0$ for $o\in\Ohat$, equals $1$ for $o\in H\setminus\Ohat$, and equals $2$ for $o\in N$. Since $T$ is the minimum of $h$ independent uniform arrival times, $\Prb(T>t)=(1-t)^h$, so $T$ has density $h(1-t)^{h-1}$ and $\E[(1-T)^k]=h/(h+k)$ for $k=1,2$. In particular, $\E[T]=1-\E[1-T]=1/(h+1)$. Averaging the bound of \Cref{lem:matroid-layers} over $T$ gives
\[
\E[v(A\cap O)+\OPT(R)]\ge v(O\cap\Ohat)+\frac{h}{h+1}v((O\setminus\Ohat)\cap H)+\frac{h}{h+2}v((O\setminus\Ohat)\setminus H).
\]
Next, an item $\hat o\in(\Ohat\setminus O)\setminus H$ belongs to $A$ exactly when $t_{\hat o}<T$, which has probability $\E[T]=1/(h+1)$, and no other item of $\Ohat\setminus O$ belongs to $A$. Thus $\E[v(A\setminus O)]=v((\Ohat\setminus O)\setminus H)/(h+1)$. Adding this to the display above and using $P=v(A\cap O)+v(A\setminus O)$ proves \eqref{eq:matroid-prefix-bound}.
\end{proof}

\begin{proof}[Proof of \Cref{thm:matroid-weighted-recovery}]
We write $P+p\OPT(R)=p\bigl(P+\OPT(R)\bigr)+(1-p)P$. As shown in the proof of \Cref{lem:matroid-prefix}, $\E[P]=v(\Ohat\setminus H)/(h+1)$. Combining this with \Cref{lem:matroid-prefix}, and splitting $\Ohat\setminus H$ into its parts inside and outside $O$, we obtain
\begin{align}
\E[P+p\OPT(R)]\ge{}&
p\,v(O\cap\Ohat\cap H)
+\left(p+\frac{1-p}{h+1}\right)v((O\cap\Ohat)\setminus H)\notag\\
&+\frac{ph}{h+1}v((O\setminus\Ohat)\cap H)
+\frac{ph}{h+2}v((O\setminus\Ohat)\setminus H)\notag\\
&+\frac1{h+1}v((\Ohat\setminus O)\setminus H).
\label{eq:matroid-weighted-coefficients}
\end{align}
The coefficient of $v((\Ohat\setminus O)\setminus H)$ is $p/(h+1)+(1-p)/(h+1)=1/(h+1)$, and the coefficient of $v((O\cap\Ohat)\setminus H)$ receives the additional term $(1-p)/(h+1)$ from the expected prefix value.

If $h\ge2$, each of the first four coefficients is at least $p/2$, since the smallest is $ph/(h+2)\ge p/2$. The last term is nonnegative, so the theorem follows from $v(O)=\OPT$.

Now suppose $H=\{f\}$. The first three coefficients in \eqref{eq:matroid-weighted-coefficients} are at least $p/2$, so the only coefficients that fall short are those of the non-violating elements of $O\setminus\Ohat$, which equal $p/3$. By the exchange property in the preliminaries applied to $\Ohat$ and $O$, there is a bijection $\phi:O\setminus\Ohat\to\Ohat\setminus O$ such that $(\Ohat\setminus\{\phi(o)\})\cup\{o\}$ is a basis for every $o\in O\setminus\Ohat$. Since $\Ohat$ maximizes the predicted value, $\vhat_o\le\vhat_{\phi(o)}$ for every such $o$.

Let $o\in O\setminus\Ohat$ be non-violating, and suppose that $\phi(o)\neq f$, so that $\phi(o)$ is also non-violating. Then
\[
(1-\theta)v_o\le\vhat_o\le\vhat_{\phi(o)}\le(1+\theta)v_{\phi(o)}.
\]
With $p=1/(1+e)$ and $\theta=(2e+1)/(2e+3)$ we have $(1-\theta)/(1+\theta)=p/2$, so $v_{\phi(o)}\ge\frac p2v_o$. The partner $\phi(o)$ lies in $(\Ohat\setminus O)\setminus H$ and has coefficient $1/2$ in \eqref{eq:matroid-weighted-coefficients}. The two elements together contribute at least
\[
\frac p3v_o+\frac12v_{\phi(o)}\ge\left(\frac p3+\frac p4\right)v_o\ge\frac p2v_o.
\]
Since $\phi$ is a bijection, no partner is used for two different elements, and the contributions of partners are disjoint from those of elements of $O$. This covers every non-violating $o\in O\setminus\Ohat$ unless $f\in\Ohat\setminus O$ and $o=o^*\defeq\phi^{-1}(f)$.

It remains to treat $f\in\Ohat\setminus O$ and the element $o^*$. Here we improve the coefficient of $v_{o^*}$ by changing the reference basis in \Cref{lem:matroid-layers}. By the exchange property, $\widetilde B=(\Ohat\setminus\{f\})\cup\{o^*\}$ is a basis, and it contains $\Ohat\setminus H=\Ohat\setminus\{f\}$. The sets $A$, $D$ and $R$, and hence the algorithm, are unchanged; only the analysis uses $\widetilde B$. Since $f\notin O$, no element of $O$ is violating. We have $a_{\widetilde B}(o)=0$ for $o\in O\cap\Ohat$; $a_{\widetilde B}(o^*)=1$, because $o^*\in\widetilde B\cap N$; and $a_{\widetilde B}(o)=2$ for every other $o\in O\setminus\Ohat$. For $h=1$, $T$ is uniform on $[0,1]$, so $\E[1-T]=1/2$ and $\E[(1-T)^2]=1/3$. Averaging \Cref{lem:matroid-layers} over $T$ as in the proof of \Cref{lem:matroid-prefix} therefore shows that \eqref{eq:matroid-prefix-bound} holds with the coefficient of $v_{o^*}$ equal to $1/2$ instead of $1/3$, and with all other coefficients unchanged. In \eqref{eq:matroid-weighted-coefficients}, the coefficient of $v_{o^*}$ thus becomes $p/2$. Every other element of $O\setminus\Ohat$ has a non-violating partner and is covered by the previous paragraph. Hence the total weight on every element of $O$ is at least $p/2$, which completes the proof.
\end{proof}

\subsection{Fractional Knapsack Secretary}

We next show that \eqref{eq:abstract-weighted-recovery} holds for the fractional knapsack secretary problem (\Cref{prob:knapsack}).

\smallskip
\noindent\textbf{Problem-Specific Notation.} Let $\xhat$ be the predicted optimal allocation and let $\bv x^*$ be a true optimal allocation, so $\OPT=\bv v\cdot\bv x^*$. Let $H$ be the set of violating items (\Cref{def:violation}), let $G=[n]\setminus H$, and let $h=|H|$. We consider instances with $h\ge1$. Let $f$ be the first violating item to arrive and let $T=t_f$. Before time $T$, the algorithm sees only items of $G$, and it accepts the fraction $\xhat_i$ of each. The value of the prefix and the capacity it consumes are therefore
\[
P=\sum_{i\in G:\,t_i<T}v_i\xhat_i\qquad\text{and}\qquad W=\sum_{i\in G:\,t_i<T}w_i\xhat_i.
\]
The residual instance $R$ consists of $f$ and all later items, and the remaining capacity is $C=1-W$. We write $\OPT(R,C)$ for the maximum value of a fractional allocation of the items in $R$ that uses capacity at most $C$. At $f$, the algorithm accepts the fraction $\min\{1,C/w_f\}$ of $f$ with probability $p$, and rejects $f$ otherwise.

\begin{theorem}\label{thm:fractional-weighted-recovery}
For any input instance to \Cref{prob:knapsack} with at least one violating item,
\begin{equation*}
  \E[P+p\OPT(R,C)]\ge\frac p2\OPT.
\end{equation*}
\end{theorem}

\begin{proof}
Let $\omega=\sum_{i\in G}w_i\xhat_i$ be the capacity used by $\xhat$ on $G$, let $\ell=\sum_{i\in G}v_ix_i^*$ be the value that $\bv x^*$ obtains from $G$, and let $u=\sum_{i\in G}v_i\xhat_i$ be the true value that $\xhat$ obtains from $G$. Let $a=\ell/\OPT$. Since $\xhat$ and $\bv x^*$ are feasible, $\omega\in[0,1]$ and $a\in[0,1]$.

We first lower bound $u$. The restriction of $\xhat$ to $G$ maximizes the predicted value among allocations of the items in $G$ with total weight at most $\omega$, since otherwise we could improve $\xhat$ by replacing this restriction and keeping its values on $H$. The allocation $\omega\cdot\bv x^*_G$ is feasible for this smaller capacity, because $\bv x^*_G$ has total weight at most $1$. Combining this with the prediction bounds on $G$, we get
\begin{equation}\label{eq:fractional-nonviolating-value}
u\ge\frac1{1+\theta}\sum_{i\in G}\vhat_i\xhat_i\ge\frac\omega{1+\theta}\sum_{i\in G}\vhat_ix_i^*\ge\frac{1-\theta}{1+\theta}\,\omega \ell=\frac{p\omega \ell}2.
\end{equation}

Now condition on $T=t$. Since $T$ depends only on the arrival times of items in $H$, each item of $G$ arrives before time $t$ independently with probability $t$. Hence $\E[P\mid T=t]=tu\ge tp\omega \ell/2$ by \eqref{eq:fractional-nonviolating-value}. For the residual value, let $L=\sum_{i\in G:\,t_i<t}v_ix_i^*$ be the value that $\bv x^*$ obtains from items that arrive before $t$. The restriction of $\bv x^*$ to the items of $R$ has value $\OPT-L$ and total weight at most $1$, so scaling it by $C=1-W$ gives an allocation that is feasible in $R$ with capacity $C$. Therefore $\OPT(R,C)\ge(1-W)(\OPT-L)$. The variables $W$ and $L$ are weighted sums of the same independent indicators $\one_{\{t_i<t\}}$ for $i\in G$, each with mean $t$. This gives $\E[W\mid T=t]=t\omega$ and $\E[L\mid T=t]=t\ell$. Moreover, $\E[WL\mid T=t]\ge t^2\omega \ell$, since the product of two distinct indicators has mean $t^2$ and the product of an indicator with itself has mean $t\ge t^2$. It follows that
\[
\E[\OPT(R,C)\mid T=t]\ge(1-t\omega)(\OPT-t\ell).
\]
Combining the bounds on prefix and residual values,
\begin{align}\label{eq:conditionT}
 \E[P + p\OPT(R,C)\mid T=t] &\ge t\,\frac{p\omega \ell}{2} + p(1-t\omega)(\OPT-t\ell)\nonumber \\
 & = p \OPT + t \left (\frac{p \omega \ell}{2} - p\omega \OPT - p\ell \right ) + t^2 p\omega \ell.
\end{align}

It remains to average over $T$. Since $T$ is the minimum of $h$ independent uniform arrival times, it has density $h(1-t)^{h-1}$, so $\E[T]=1/(h+1)$ and $\E[T^2]=2/((h+1)(h+2))$. Substituting these into \eqref{eq:conditionT} and dividing by $\OPT$ gives
\[
\frac{\E[P + p\OPT(R,C)]}{\OPT} \ge p - \frac{p(\omega+a)}{h+1} + \frac{p\omega a}{2(h+1)} + \frac{2p\omega a}{(h+1)(h+2)}.
\]
The right side is bilinear in $(\omega,a)$, so over $[0,1]^2$ its minimum is attained at a corner. The values at the corners $(0,0)$, $(0,1)$, $(1,0)$ and $(1,1)$ are
\[
p,\qquad \frac{ph}{h+1},\qquad \frac{ph}{h+1},\qquad \frac{ph}{h+2}+\frac{p}{2(h+1)}.
\]
Each is at least $p/2$. This is clear for the first three since $h\ge1$. For the last, $h/(h+2)\ge1/2$ if $h\ge2$, and for $h=1$ the value is $7p/12$. Hence $\E[P+p\OPT(R,C)]\ge\frac p2\OPT$.
\end{proof}

\subsection{Bipartite Matching}

Finally, we show that \eqref{eq:abstract-weighted-recovery} holds for online bipartite matching with random arrivals (\Cref{prob:matching}).

\smallskip
\noindent\textbf{Problem-Specific Notation.} We write each edge as $e=(i,u)$, with online endpoint $i\in[n]$ and offline endpoint $u\in U$. Let $\Mhat$ and $M^*$ be the matchings of maximum predicted and true value, and for a set of edges $A$ write $v(A)=\sum_{e\in A}v_e$ and $\vhat(A)=\sum_{e\in A}\vhat_e$. Let $H$ be the set of violating vertices (\Cref{def:violation}) and let $h=|H|\ge1$. Let $f$ be the first violating vertex to arrive and let $T=t_f$. Before time $T$, whenever a vertex arrives, the algorithm adds its edge in $\Mhat$ if it has one. Let $P$ be the total true value of these edges. The residual instance $R$ consists of $f$ and all later vertices, together with the offline vertices that are not matched before $T$.

\begin{theorem}\label{thm:matching-weighted-recovery}
For any input instance to \Cref{prob:matching} with at least one violating vertex,
\begin{equation*}
  \E[P+p\OPT(R)]\ge\frac p2\OPT.
\end{equation*}
\end{theorem}

\begin{proof}
Say that an edge $e=(i,u)\in M^*$ survives if $i$ is in $R$ and $u$ is not matched before $T$. The surviving edges of $M^*$ form a matching in $R$, so $\OPT(R)$ is at least their total value. Let $B_e$ consist of $i$ if $i\notin H$, together with the online endpoint $i'$ of the edge of $\Mhat$ at $u$ if this edge exists and $i'\notin H$. Then $e$ survives exactly when every vertex of $B_e$ arrives after $T$, that is, when the first vertex to arrive among $H\cup B_e$ belongs to $H$. The sets $H$ and $B_e$ are disjoint and $|B_e|\le2$, so
\begin{equation}\label{eq:matching-survival}
\Prb(e\text{ survives})=\frac{h}{h+|B_e|}\ge\frac{h}{h+2}.
\end{equation}

If $h\ge2$, this probability is at least $1/2$ for every edge of $M^*$. Hence $\E[\OPT(R)]\ge\OPT/2$, and since $P\ge0$ we get $\E[P+p\OPT(R)]\ge p\,\E[\OPT(R)]\ge\frac p2\OPT$.

Now suppose $H=\{f\}$. Let $\overline M$ be the set of edges of $\Mhat$ whose online endpoint is not $f$. Each edge of $\overline M$ is accepted before $f$ with probability $1/2$, so $\E[P]=\frac12v(\overline M)$. Let $Q$ be the set of edges $(i,u)\in M^*\setminus\Mhat$ with $i\ne f$ such that $u$ is matched in $\Mhat$ to a vertex $i'\ne f$. For $e\in Q$ the vertices $f$, $i$ and $i'$ are distinct, and $e$ survives exactly when $f$ arrives before both $i$ and $i'$. This has probability $1/3$. Every other edge $e\in M^*$ has $|B_e|\le1$, so it survives with probability at least $1/2$ by \eqref{eq:matching-survival}. Therefore
\[
\E[\OPT(R)]\ge\frac13v(Q)+\frac12v(M^*\setminus Q).
\]

It remains to relate $v(\overline M)$ to $v(Q)$. Let $\Gamma\subseteq\Mhat$ be the set of predicted edges incident to an endpoint of an edge in $Q$. Since $Q\subseteq M^*$ is a matching, replacing $\Gamma$ by $Q$ in $\Mhat$ gives a matching, and predicted optimality gives $\vhat(\Gamma)\ge\vhat(Q)$. Every edge in $Q\cup\Gamma$ has an online endpoint other than $f$: this holds by definition for edges of $Q$, for predicted edges at the online endpoint of an edge of $Q$, and, by the definition of $Q$, for predicted edges at its offline endpoint. All these edges are therefore non-violating, and
\[
(1-\theta)v(Q)\le\vhat(Q)\le\vhat(\Gamma)\le(1+\theta)v(\Gamma).
\]
Since $\Gamma\subseteq\overline M$ and $(1-\theta)/(1+\theta)=p/2$, this gives $v(\overline M)\ge v(\Gamma)\ge\frac p2v(Q)$. Combining the bounds on $\E[P]$ and $\E[\OPT(R)]$,
\begin{align*}
 \E[P+p\OPT(R)]
 &\ge\frac12v(\overline M)+p\left(\frac13v(Q)+\frac12v(M^*\setminus Q)\right)\\
 &\ge\left(\frac p4+\frac p3\right)v(Q)+\frac p2v(M^*\setminus Q)\\
 &\ge\frac p2v(M^*),
\end{align*}
which proves the theorem.
\end{proof}

\section{Integral Knapsack Secretary} \label{sec:integral}

In this section we show that, despite the positive result for fractional knapsack secretary, the integral version does not admit a nontrivial consistency--robustness tradeoff. Accepting a small item can make a later item of weight one infeasible, and the value of that later item may be underestimated by the prediction. Formally, we show:

\begin{theorem}\label{thm:integral-upper}
Fix $\delta\in[0,1]$. Any algorithm for integral knapsack secretary that achieves expected value at least $(1-\delta)\OPT$ whenever all predictions are correct has worst-case competitive ratio at most $\delta$.
\end{theorem}
The bound of \Cref{thm:integral-upper} is tight up to a constant factor. Consider the algorithm that runs the $0.178$-competitive prediction-free algorithm of \citet{BalkanskiEtAl2026} with probability $\delta$, and follows the predicted optimum otherwise. It achieves expected value at least $(1-\delta)\OPT$ whenever all predictions are correct, and has worst-case competitive ratio $0.178\delta$.
\begin{proof}
The idea is to use many weight-one items as ``traps.'' On an instance with correct predictions, consistency forces the algorithm to accept small items early, so most weight-one items arrive when the knapsack is already nonempty and therefore cannot be accepted. We then make one such item very valuable. The algorithm cannot notice this change before that item arrives.

Fix an integer $k\ge 2$. There are $k$ \emph{small} items, each with weight and value $1/k$, and $k$ \emph{large} items, each with weight $1$ and value $1/k$. All predictions are correct, and $\OPT=1$, achieved by taking all small items. Let $\ALG_0$ be the algorithm's value, and let $X$ be the number of small items it rejects. A large item fits only into an empty knapsack, so at most one large item is accepted. Hence $\ALG_0\le (k-X)/k+1/k$. By consistency, $\E[\ALG_0]\ge 1-\delta$, and therefore
\begin{equation}\label{eq:integral-small-rejections}
    \E[X]\le \delta k+1.
\end{equation}

For a large item $z$, let $E_z$ be the event that the knapsack is empty when $z$ arrives, and let $N=\sum_{z\text{ large}}\one_{E_z}$. Let $L_t$ and $S_t$ be the numbers of large and small items among the first $t$ arrivals, and let $Y=\max_{0\le t\le 2k}(L_t-S_t)\ge 0$. Note that $Y$ depends only on the arrival order, not on the algorithm.

Let $s$ be the arrival position of the first accepted item, or $s=2k$ if nothing is accepted. Exactly the large items among the first $s$ arrivals see an empty knapsack, so $N=L_s$. Every small item among the first $s$ arrivals is rejected, except possibly the one at position $s$. Therefore $S_s\le X+1$, and
\[
    N=S_s+(L_s-S_s)\le X+1+Y.
\]

It remains to bound $\E[Y]$. The arrival order induces a uniformly random $\pm1$ path from $0$ to $0$ with $2k$ steps, where each large arrival is a $+1$ step and each small arrival is a $-1$ step. For $1\le r\le k$, reflecting a path after it first reaches level $r$ gives a bijection between paths that reach $r$ and all $\pm1$ paths ending at $2r$. Hence
\[
    \Prb(Y\ge r)=\frac{\binom{2k}{k+r}}{\binom{2k}{k}}
    =\prod_{j=1}^{r}\frac{k-j+1}{k+j}
    \le\prod_{j=1}^{r}\Bigl(1-\frac{2j-1}{2k}\Bigr)
    \le e^{-r^2/(2k)},
\]
so $\E[Y]=\sum_{r\ge1}\Prb(Y\ge r)\le\int_0^\infty e^{-x^2/(2k)}\,dx =\sqrt{\pi k/2}$. Combining this with~\eqref{eq:integral-small-rejections} gives
\begin{equation}\label{eq:integral-empty-average}
    \frac1k\sum_{z\text{ large}}\Prb(E_z)=\frac{\E[N]}{k}
    \le \delta+\frac2k+\sqrt{\frac{\pi}{2k}}.
\end{equation}

By~\eqref{eq:integral-empty-average}, some large item $z^*$ satisfies $\Prb(E_{z^*})\le\delta+2/k+\sqrt{\pi/(2k)}$. This item depends only on the algorithm and the accurate instance, not on the random order. Change the true value of $z^*$ to $k$, while keeping its weight $1$ and its prediction $1/k$. Use the same arrival order and the same internal randomness in both instances. Before $z^*$ arrives, the algorithm sees exactly the same input, so it makes the same decisions. In particular, $E_{z^*}$ is the same event in both runs.

On the modified instance, $\OPT_1=k$. The algorithm can accept $z^*$ only on $E_{z^*}$. Without $z^*$, its value is at most $1$. Therefore
\[
    \frac{\E[\ALG_1]}{\OPT_1}\le \Prb(E_{z^*})+\frac1k
    \le \delta+\frac3k+\sqrt{\frac{\pi}{2k}}.
\]
Letting $k\to\infty$ shows that the worst-case competitive ratio is at most $\delta$.
\end{proof}

\paragraph{AI Methodology.} We used ChatGPT and Claude to assist with the refinement of material in \Cref{sec:applications,sec:integral}. The authors verified the correctness of all content, including all mathematical arguments and references, and take full responsibility for the paper.

\bibliographystyle{plainnat}
\bibliography{reference}
\end{document}

%% file: bib_macros.tex
  \usepackage{nth}
  \usepackage{intcalc}

  \newcommand{\cAAAI}[1]{AAAI\ Conference\ on\ Artificial (AAAI)}